\documentclass[twocolumn,11pt]{article}
\usepackage{times}
\usepackage[dvips]{graphicx}

\newtheorem{definition}{Definition}

\begin{document}
\global\def\refname{{\normalsize \it References:}}
\baselineskip 12.5pt
%
%
%
\title{\hspace{-11pt}\LARGE \bf Cognitive Binary Logic - The Natural Unified Formal Theory of Propositional Binary Logic}

\date{}

\author{\hspace*{-10pt}
\begin{minipage}[t]{3.5in} \normalsize \baselineskip 12.5pt
\centerline{\hspace{-11pt} NICOLAIE POPESCU-BODORIN, \textit{Member}, \textit{IEEE}}
\centerline{\hspace{-11pt} Spiru Haret University}
\centerline{\hspace{-11pt} Dept. of Mathematics and Computer Science}
\centerline{\hspace{-11pt} 13 Ion Ghica, Bucharest 3}
\centerline{\hspace{-11pt} ROMANIA}
\centerline{\hspace{-11pt} http://fmi.spiruharet.ro/bodorin/}
\end{minipage} \kern 0in
\begin{minipage}[t]{3.5in} \normalsize \baselineskip 12.5pt
\centerline{LUMINI\c{T}A STATE}
\centerline{University of Pite\c{s}ti}
\centerline{Dept. of Mathematics and Computer Science}
\centerline{1 Targu din Vale, Pite\c{s}ti, Arge\c{s}}
\centerline{ROMANIA}
\centerline{lstate@clicknet.ro}
\end{minipage}
%
%
\\ \\ \hspace*{-15pt}
\begin{minipage}[b]{6.8in} \normalsize
\baselineskip 12.5pt {\it Abstract:}
This paper presents a formal theory which describes propositional binary logic as a semantically closed formal language, and allows for syntactically and semantically well-formed formulae, formal proofs (demonstrability in Hilbertian acception), deduction (Gentzen's view of demonstrability), CNF-ization, and deconstruction to be expressed and tested in the same (computational) formal language, using the same data structure. It is also shown here that \textit{Cognitive Binary Logic} is a self-described theory in which the \textit{Liar Paradox} is deconstructed. 
\\ [4mm] {\it Key--Words:}
cognitive binary logic, inductive\slash deductive discourse, liar paradox, computational logic
\end{minipage}
\vspace{-15pt}}

\maketitle

\thispagestyle{empty} \pagestyle{empty}
%
%
\section{Introduction}
\label{S1} \vspace{-4pt}

This paper presents a unified natural approach to propositional binary logic, in which syntactically and semantically well-formed formulae, formal proofs (demonstrability in Hilbertian acception), deduction (Gentzen's view of demonstrability, \cite{gentzen}), CNF-ization \cite{prooftheory85}, and deconstruction are expressed and tested in the same (computational) formal language - Computational Cognitive Binary Logic ($CCBL$), using the same data structure: the deductive discourse. 

It is also shown here that \textit{Cognitive Binary Logic} is a self-described theory in which the \textit{Liar Paradox} \cite{liar-paradox} is deconstructed. 

The prerequisites of this paper are the following concepts: 
\textit{formal language} \cite{formal-language}, 
\textit{formal theory} (formal system) \cite{formal-theory}, 
\textit{formal proof} \cite{formal-proof}, 
\textit{deduction} \cite{gentzen},  
\textit{resolution} \cite{prooftheory85}, 
\textit{valid argumentation} \cite{formal-proof}, 
\textit{Gentzen's natural deduction system} (sequent calculus) \cite{gentzen}, 
\textit{completeness} \cite{formal-proof}, 
\textit{consistency} (soundness) \cite{formal-proof}  - all of them considered in the context of propositional binary logic, 
\textit{equivalence relations}, 
\textit{modal logic} \cite{ML}, 
\textit{Lukasiewicz's} \cite{Lukasiewicz} and 
\textit{Hilbert's} \cite{Hilbert} classical formalizations of binary logic. 
Since they are so many and the space here is limited, we would like to refer to bibliographic sources instead of reproducing  some redundant contents. 

Despite the vast universe of discourse, this paper is based on a single assumption which we all know to be true: both Lukasiewicz's and Hilbert's formalizations of propositional binary logic are complete and sound formal theories. 


\subsection{Outline}
\vspace{-4pt}

The structure of this paper follows an imaginary process of reverse engineering the Propositional Binary Logic ($BPL$). We start by analyzing how it works in order to see what is it made of. 

Section 2 discuss the prerequisites of the proposed formal theory. Cognitive implication is introduced here as a natural way of modelling human logical thinking as opposite to the trivial sintactic truth \textit{ex contradictione quodlibet}.

A dual Semantic-Computational formalization of Cognitive Binary Logic ($SCBL, CCBL$) is presented in the third section of the paper. It is shown here that both $SCBL$ and $CCBL$ theories cover $BPL$ in a complete and consistent manner. Also, in $CCBL$ the inductive/deductive proof of any theorem is algoritmically computable. In the end, it is shown that even if $CCBL$ theory qualify the language of $PBL$ as being semantically closed, the Liar Paradox is still successfully deconstructed.

\vspace{-8pt}
\section{Setting the semantic framework}
\label{S2} \vspace{-4pt}

From the beginning we must say that present computers can only parse and evaluate data structures that are truly meaningful only in human understanding. 

Hence, our world is a semantic one, while the computational world is defined, regulated, controlled and described through syntactic rules. But even when we think about syntactic rules, we still operate at the semantic level for the simple reason that symbols themselves are not very important to us, while their meaning is our actual concern. 

\subsection{Formalization}
\label{S2-1} \vspace{-4pt}

The symbols become important only when it comes to formulate our semantic knowledge into something computable by expressing our understanding into a computational language. In case of binary logic, the challenge is to translate human reasoning into something a little bit more sophisticated than (two-element) Boolean algebra while achieving a  certain degree of semantic fidelity measured in terms of similitude or analogy with human thinking. This is what we usually call \textit{formalization} and the result is a \textit{formal theory} defined over a \textit{formal language}. 

Since we cannot formalize something unless we are fully aware of its meanings, a formalization will never come easy because it requires a triple effort.

The first step is to understand the world that is to be formalized, and this is what we do here, throughout the second section.

The second step is to associate  semantic charges (the meanings) to the assemblies of symbols belonging to a pure syntactic world while setting it up.

Last, but not least, we must test the semantic efficiency of the newly created formal system by verifying its limits in terms of completeness and consistency.

The second step is particularly difficult, and probably this is the reason why the syntactic rules (the ways of constructing well-formed formulae) and the reasoning rules (the ways of constructing well-formed proofs) are very different in the present formal theories of the propositional binary logic. 

\subsection{Implication and human reasoning}
\label{S2-2} \vspace{-4pt}

Let us comment on a well-known example: \textit{I think, therefore I am}. 
It is inevitably true that I am, since I know that I think and, on the other hand, I also know that if I did not exist, then certainly I would not think.

Now, let us consider that $p$ and $q$ are propositional variables with $q$ being a tautology and $p$ being logically false. Hence, $a=(p \rightarrow q)$ is true but still, in human understanding $a$ is not a valid argument for $q$. Syntactically, an  implication can be parsed and evaluated as true, but this does not necessarily qualify that implication as a valid argument. 

It can be seen in the above examples that a valid argument in human logical thinking is a true implication with true premise. Humans think semantically, not syntactically, by using some implications as valid arguments. By the way, nobody has said: \textit{I don't doubt or I think, I don't think or I am}.

\subsection{One more challenge} 
\label{S2-3} \vspace{-4pt}

We should ask ourselves why the following truth is told surprisingly rarely: \textit{the natural logical framework for analyzing the formulae of propositional binary logic language is modal logic} \cite{ML}. 

The above assertion is true because any formula within the language falls into one of the following three categories:  formulae which are \textit{always true} (tautologies), formulae which are \textit{sometimes true} (and sometimes false) and formulae which are \textit{always false}. By reformulating in classical terminology of modal logic, a formula could be: \textit{necessary} true (an axiom or a theorem, or, generally speaking, a tautology) or \textit{possibly} true (a contingent, or a formula demonstrable under some hypotheses which give \textit{the context} of that \textit{satisfiable} truth) or \textit{impossibly} true (i.e. \textit{necessary} false, or contradiction).

If $FORM$ denotes the set of formulae within the language of propositional binary logic, then: $FORM=\hat{f} \cup \hat{c_{t}} \cup \hat{t}$, where $\hat{f}$ is the class of necessary false formulae, $\hat{c_{t}}$ is the class of contextual truths (or the class of proper variables whose truth values are not constant) and $\hat{t}$ is the class of all tautologies.

Hence, one more challenge here is to put this three-valued state of truth in the binary framework of the propositional binary logic and to explain what makes this possible.

\subsection{Cognitive implication}
\label{S2-4} \vspace{-4pt}

A cognitive implication is assumed here to be a formula of the following type: 
\begin{eqnarray}\label{ec-cognitive-implication}
\left( \bigwedge_{i=1}^{n}h_{i} \right) \rightarrow \left( \bigvee_{i=1}^{m}c_{j} \right),
\end{eqnarray}
where $\{ h_{i} \}_{1 \leq i \leq n}$ and $\{ c_{j} \}_{1 \leq j \leq m}$ are two sets of formulae (hypotheses and conclusions, respectively).

The name is suggested by the fact that investigating a universe is a matter of finding a suitable collection of hypotheses assumed to be simultaneously true (hence, their conjuntion is true) and studying the possibilities to infer some conclusions which are not necessarily simultaneously true. To demonstrate how suitable this structure is for describing human cognitive processes, we give the following example: we want to build a formal theory over a formal language. But what does that mean exactly? Nothing more than identifying a set of simoultaneously true premises (axioms and argumentation rules) that enable us to prove or to disprove well-formed sentences (formulae) written in that language. 

\subsection{Implication as a SAT problem}
\label{S2-5} \vspace{-4pt}

A trivial cognitive implication in propositional binary logic is that no matter the hypothesis $h$, the \textit{tertium non datur} is always true and so it is the implication:
\begin{eqnarray}\label{ec-trivial-cognitive-implication1}
h \rightarrow \left( a \vee \neg a \right). 
\end{eqnarray}
Nothing changes if \textit{tertium non datur} is replaced by any other tautology $t$: 
\begin{eqnarray}\label{ec-trivial-cognitive-implication2}
h \rightarrow t, 
\end{eqnarray}
Also, no matter the conclusion $c$, if the hypotesis is a contradiction, the following implication is true:
\begin{eqnarray}\label{ec-trivial-cognitive-implication3}
f \rightarrow c. 
\end{eqnarray}
Hence, there is no doubt that formulae 
(\ref{ec-trivial-cognitive-implication1})-(\ref{ec-trivial-cognitive-implication3}) 
are trivial Boolean satisfiability (SAT) problems. 

Since formula (\ref{ec-trivial-cognitive-implication3}) is by default a tautology, it follows that \textit{there is only one type of implication that deserves to be studied in binary logic}: 
\begin{eqnarray}\label{ec-single-non-trivial-implication}
s_{t} \rightarrow  c, 
\end{eqnarray}
where $s_{t} \in \hat{s}_{t}$ and:
\begin{eqnarray}\label{ec-single-non-trivial-implication}
\hat{s}_{t} = \hat{c}_{t} \cup \hat{t}, 
\end{eqnarray} 

\subsection{SAT problem vs. cognitive implication}
\label{S2-6} \vspace{-4pt}

Let $a \in FORM= \hat{f} \cup \hat{c}_{t} \cup \hat{t}$ an arbitrary formula. Its corresponding SAT problem can be stated as a decision problem as follows:
\begin{eqnarray}\label{ec-SAT}
 (a \in \hat{t}) \oplus (a \in \hat{c}_{t}) \oplus (a \in \hat{f}),
\end{eqnarray} 
where $\oplus$ denotes exclusive disjunction. On the other hand, 
$\forall (a,f,t) \in FORM \times \hat{f} \times \hat{t} $:
\begin{eqnarray}\label{ec-any-formula-is-an-implication4}
		\left[
			(t \rightarrow (a \vee f))
			\leftrightarrow
			a
		\right] 
		\in \hat{t}
\end{eqnarray}

In other words, the formulae (\ref{ec-SAT}), (\ref{ec-any-formula-is-an-implication4}) tell us that the demonstration of \textit{the truth} or of a \textit{contextual truth}, and also the \textit{resolution}  of a contextual truth (the process of finding a context which makes a formula satisfiable), all of them can be carried out by parsing the same data structure, which is syntactically a  cognitive implication and, semantically, is an attemp to find a valid argument for that truth. 

Also, all three modal states of truth ($\hat{t},\hat{f},\hat{c_{t}}$) are decidable in propositional binary logic using the same test: a cognitive implication (\ref{ec-any-formula-is-an-implication4}).

\section{Cognitive Binary Logic}
\label{S3} \vspace{-4pt}

A dual formalization of Cognitive Binary Logic ($CBL$) is defined here (N. Popescu-Bodorin) as a pair of formal theories given over the same language (propositional binary logic), using the same set of axioms (\textit{tertium non datur}) and two different sets of valid arguments which still produce the same set of theorems. 

\subsection{The axiom}\label{sbs_the_axiom}
\label{S3-1} \vspace{-4pt}

There is a single axiom in $CBL$: \textit{tertium non datur}, considered in the classical form:
\begin{eqnarray}\label{ec-the-axiom1}
a \vee \neg a, 
\end{eqnarray}
or in the simplest form of a cognitive implication:
\begin{eqnarray}\label{ec-the-axiom2}
a \rightarrow a, 
\end{eqnarray}
or in a more general form:
\begin{eqnarray}\label{ec-the-axiom3}
\left[ a \wedge \left( \bigwedge_{i=1}^{n}h_{i} \right) 
\right]
\rightarrow 
\left[ a \vee \left( \bigvee_{i=1}^{m}c_{j} \right)
\right].
\end{eqnarray}\\
where the meanings of symbols appearing in (\ref{ec-the-axiom1})-(\ref{ec-the-axiom3}) are already introduced in the previous subsections.

\subsection{Semantic formalization of CBL}
\label{S3-2} \vspace{-4pt}

The first set of valid arguments used here to define a first formalization of $CBL$ is meaningful in human understanding. It is further denoted as $\mathcal{A}_{1}$ and contains the following equivalences:\\
\begin{enumerate}
\item The law of double negation:
\begin{eqnarray}\label{ec-varguments-1-1}
 a \Leftrightarrow \neg \neg a.
\end{eqnarray}
\item The law of the contrapositive:
\begin{eqnarray}\label{ec-varguments-1-2}
\left(a \rightarrow b \right) \Leftrightarrow \left(\neg b \rightarrow \neg a \right).
\end{eqnarray}
\item The law of deduction-resolution:
\begin{eqnarray}\label{ec-varguments-1-3}
\left[\left( h \wedge a \right) \rightarrow b \right]
\Leftrightarrow
\left[ h \rightarrow \left( a \rightarrow b \right) \right].
\end{eqnarray}
\item Distributivity law:
$$\left\lbrace 
	h
	\rightarrow
	\left[ 
		c
		\vee
		\left ( 
			a \wedge b
		\right )
	\right] 
\right\rbrace  
\Leftrightarrow 
$$
\begin{eqnarray}\label{ec-varguments-1-4}
\Leftrightarrow 
\left\lbrace 
	\left[ 
		h 
		\rightarrow 
		\left( 
			c \vee a 
		\right) 
	\right] 
	\wedge 
	\left [ 
		h
		\rightarrow 
		\left(  
			c
			\vee
			b
		\right)
	\right ] 
\right\rbrace. 
\end{eqnarray}
\end{enumerate}
Let us denote the semantic formalization of the $CBL$ as being the triplet formed with the language of propositional logic ($L$), \textit{tertium non datur} axiom ($TND$) and the set of valid arguments $\mathcal{A}_{1}$ from above: 
\begin{eqnarray}\label{ec-LBL-def}
SCBL=\{L, TND, \mathcal{A}_{1}\}. 
\end{eqnarray}

Let $LBL$ be Lukasiewicz's classical formalization of propositional binary logic: 
\begin{eqnarray}\label{ec-LBL-def}
LBL=\{L, Axiom, MP\}, 
\end{eqnarray}
where $L$ denotes the formal language of propositional binary logic, $MP$ stands for \textit{Modus Ponens}, and $Axiom$ is the following set of axioms:
\begin{eqnarray}\label{ec-LBL-axiom-1}
 p \rightarrow (q \rightarrow p), 
\end{eqnarray}
\begin{eqnarray}\label{ec-LBL-axiom-2}
 [p \rightarrow (q \rightarrow r)] \rightarrow [(p \rightarrow q) \rightarrow (p \rightarrow r)],
\end{eqnarray}
\begin{eqnarray}\label{ec-LBL-axiom-3}
 (\neg p \rightarrow \neg q) \rightarrow (q \rightarrow p). 
\end{eqnarray}
Since $TND$ and the formulae of $\mathcal{A}_{1}$ are theorems in $LBL$, it is clear that all theorems of $SCBL$ will be  theorems in $LBL$, i.e. $SCBL$ is a sub-theory of $LBL$, fact that will be denoted as: 
\begin{eqnarray}\label{ec-SCBL<LBL}
SCBL \subset LBL.
\end{eqnarray}

$SCBL$ is an inductive (Hilbertian) theory. Writting a formal proof in $SCBL$ is a matter of experience and intuition. We `guess' what axioms and what substitutions we must use in order to prove a theorem. This approach is not suitable for algorithmic computation because from this point of view, `guessing' means an exhaustive search in an infinite set of axiom instances. Still, the formulae within the second set of valid arguments ($\mathcal{A}_{2}$, given below) are very easy to prove using $SCBL$ formalization.

In the computational formalization of the $CBL$, we aim to make any formal proof algorithmically computable. The general ideea is to use formula \ref{ec-any-formula-is-an-implication4}, i.e. the fact that any formula $a$ is equivalent to a cognitive implication $(t \rightarrow (a \vee f))$, which is further decomposable as a conjunction of cognitive implications. Therefore, the set of valid arguments will be appropriate for this purpose, more redundant, and less intelligible at a first glance.

\subsection{Computational formalization of CBL}
\label{S3-3} \vspace{-4pt}

Computational formalization of Cognitive Binary Logic ($CCBL$) is defined by a second set of valid arguments, denoted $\mathcal{A}_{2}$, any argument being an equivalence and also a rule of syntactic and semantic simplification\slash complexification (elimination\slash introduction rule for a logical connective):

\begin{enumerate} 
\item First distributivity law (Right-side conjunction elimination / introduction rule): 
$$\left\lbrace
	h
	\rightarrow
	\left[ 
		c
		\vee
		\left ( 
			\alpha \wedge \beta
		\right )
	\right] 
\right\rbrace  
\Leftrightarrow $$
\begin{eqnarray}\label{ec-varguments-2-1}
\Leftrightarrow
\left\lbrace 
	\left[ 
		h 
		\rightarrow 
		\left( 
			c \vee \alpha 
		\right) 
	\right] 
	\wedge 
	\left [ 
		h
		\rightarrow 
		\left(  
			c
			\vee
			\beta
		\right)
	\right ] 
\right\rbrace. 
\end{eqnarray}
\item Second distributivity law (Left-side disjunction elimination / introduction rule):  
$$\left\lbrace 
	\left[
		h
		\wedge
		\left(
			\alpha \vee \beta
		\right)
	\right]
	\rightarrow
	c
\right\rbrace 
\Leftrightarrow $$
\begin{eqnarray}\label{ec-varguments-2-2}
\Leftrightarrow
\left\lbrace 
	\left[
		\left(
			h
			\wedge
			\alpha
		\right)
		\rightarrow
		c
	\right]
	\wedge
	\left[
		\left(
			h
			\wedge
			\beta
		\right)
		\rightarrow
		c
	\right]
\right\rbrace. 
\end{eqnarray}
\item First reformulation of deduction-resolution law (Left-side negation elimination / introduction rule): 
\begin{eqnarray}\label{ec-varguments-2-3}
\left\lbrace 
	\left[
		h
		\wedge
		\left(
			\neg \alpha
		\right)
	\right]
	\rightarrow
	c
\right\rbrace 
\Leftrightarrow 
\left\lbrace 
	h
	\rightarrow
	\left[
		c
		\vee
		\alpha
	\right]
\right\rbrace. 
\end{eqnarray}
\item Second reformulation of deduction-resolution law (Right-side negation elimination / introduction rule): 
\begin{eqnarray}\label{ec-varguments-2-4}
\left\lbrace 
	h
	\rightarrow
	\left[
		c
		\vee
		(\neg \alpha)
	\right]
\right\rbrace 
\Leftrightarrow 
\left\lbrace 
	\left[
		h
		\wedge
		\alpha
	\right]
	\rightarrow
	c
\right\rbrace. 
\end{eqnarray}
\item Third reformulation of deduction-resolution law (Left-side implication elimination / introduction rule): 
$$\left\lbrace 
	\left[
		h
		\wedge
		\left(
			\alpha \rightarrow \beta
		\right)
	\right]
	\rightarrow
	c
\right\rbrace 
\Leftrightarrow $$
\begin{eqnarray}\label{ec-varguments-2-5}
\Leftrightarrow
\left\lbrace 
	\left[
		\left(
			h
			\wedge
			\beta
		\right)
		\rightarrow
		c
	\right]
	\wedge
	\left[
		h
		\rightarrow
		\left(
			c
			\vee
			\alpha
		\right)
	\right]
\right\rbrace. 
\end{eqnarray}
\item Generalized deduction-resolution law (Right-side implication elimination / introduction rule): 
$$\left\lbrace 
	h
	\rightarrow
	\left[
		c
		\vee
		\left(
			\alpha \rightarrow \beta
		\right)
	\right]
\right\rbrace 
\Leftrightarrow $$
\begin{eqnarray}\label{ec-varguments-2-6}
\Leftrightarrow
\left\lbrace 
	\left[
		h
		\wedge
		\alpha
	\right]
	\rightarrow
	\left[
		c
		\vee
		\beta
	\right]
\right\rbrace. 
\end{eqnarray}
\end{enumerate}

The formulae within $\mathcal{A}_{2}$ are theorems of $SCBL$, hence we get: 
\begin{eqnarray}\label{ec-CCBL<SCBL<LBL}
CCBL \subset SCBL \subset LBL.
\end{eqnarray}

\begin{definition}
\textit{CCBL Formal Theory} (N. Popescu-Bodorin): \\
Let $L=(B,V, C, S, G_{L}, FORM) $ be the classical formal language of propositional binary logic, and let $LBL=\{L, Axiom, MP\}$ be the Lukasiewicz formalization of propositional binary logic, where: 
\begin{itemize}
\item\vspace{-5.75pt} 
$B=\{0,1\}$ is the set of binary truth values;
\item\vspace{-5.75pt} 
$V$ is the set of propositional variables; 
\item\vspace{-5.75pt} 
$C$ is the collection of logical connectives, $C=\{\neg, \wedge, \vee, \rightarrow, \leftrightarrow \}$, which are all defined using truth tables (logical gates); 
\item\vspace{-5.75pt} 
$S=\{(,)\}$ is the list of separators; 
\item\vspace{-5.75pt} 
$G_{L}$ is the formal grammar which qualifies the strings from $V \cup C \cup S$ as well-formed formulae (the elements of $FORM$, i.e. syntactically valid sentences) through recursive structural complexification with logical connectives from $C$ and symbols (propositional variables) from $V$.
\item\vspace{-5.75pt} 
$FORM=\hat{t} \cup \hat{c_{t}} \cup \hat{f}$ is the set of all well-formed formulae classified by their modal state of truth: tautologies, contextual truths and contradictions.
\end{itemize}

Let $\mathcal{L}=(\mathcal{B},\mathcal{R}, \mathcal{V}, C, S, \mathcal{G}_{\mathcal{L}}, CI, \mathcal{D}, D) $ be the formal language of $CCBL$, where: 
\begin{eqnarray}\label{ec-CCBL-def}
CCBL=\{\mathcal{L}, TND, \mathcal{A}_{2}\},
\vspace{-5.75pt}
\end{eqnarray}
and:
\begin{itemize}
\item\vspace{-5.75pt}
$\mathcal{A}_{2}$ is the set of valid arguments (\ref{ec-varguments-2-1})-(\ref{ec-varguments-2-6});
\item\vspace{-5.75pt}
$TND$ stands for \textit{tertium non datur} (\ref{ec-the-axiom1})-(\ref{ec-the-axiom3});
\item\vspace{-5.75pt}
$\mathcal{B}=\{\hat{t}, \hat{c_{t}}, \hat{f}\}$, i.e. the discourse in $\mathcal{L}$ is focused on tautologies, contextual truths and contradictions;
\item\vspace{-5.75pt}
$\mathcal{R}=\{t, c_{t}, f\}$ is the list of reserved symbols used to refer a generic variable from $\hat{t}$, $ \hat{c_{t}}$ and $ \hat{f}$, respectively;
\item\vspace{-5.75pt}
$\mathcal{V}=FORM$, i.e. the discourse in $\mathcal{L}$ is a meta-discourse on $L$ asserting something about the formulae within $L$; 
\item\vspace{-5.75pt}
$\mathcal{G}_{\mathcal{L}}$ is the formal grammar that qualifies the cognitive implications (units of discourse, like simple sentences in natural language) as the simplest well-formed formulae within $\mathcal{L}$ using a single rule of immersion of $FORM$ in $CI$: 
\vspace{-5.25pt}
$$\{\alpha \in FORM\} \Leftrightarrow$$
\vspace{-20pt}
\begin{eqnarray}\label{ec-CCBL-gramar}
\Leftrightarrow \{\alpha_{ci}=\left[ t \rightarrow (\alpha \vee f) \right] \in CI \}.
\vspace{-5.25pt}
\end{eqnarray}
$\mathcal{G}_{\mathcal{L}}$ also recursively qualifies more complex structure of discourse (like complex sentences in natural language) as being well-formed by using six rules of phrasal grammar, namely formulae (\ref{ec-varguments-2-1})-(\ref{ec-varguments-2-6}) from the set $\mathcal{A}_{2}$; 
\item\vspace{-5.75pt}
$CI$ contains the simplest formulae of the language which are cognitive implications and conjunctions of cognitive implications; 
\item\vspace{-5.75pt}
The \textit{full deductive discourse} about a cognitive implication $\alpha_{ci}$ within $\mathcal{L}$ is defined as a tree having the following properties:\\
- The root is labeled with $\alpha_{ci}$, or else, $\alpha$ is the label of the root and the expansion rule to be applied is the immersion law (\ref{ec-CCBL-gramar}) and, consequently, the outer degree of the root is 1 and the only descendant of the root is labeled with $\alpha_{ci}$; \\
- Starting with the vertex labeled with $\alpha_{ci}$, tree expansion is conducted by the rules of $\mathcal{A}_{2}$ (applied from the left to the right), up to the point where they are no longer applicable;\\
Any partial expansion is simply called a \textit{deductive discourse} about $\alpha_{ci}$. We denote by $\mathcal{D}$, the set of all possibles deductive discourses about the formulae within $\mathcal{V}=FORM$. 
\item\vspace{-5.75pt}
The summary of a full deductive discourse (with $n$ leaves) expanded for a formula $a \in FORM$ can be written as:
\begin{eqnarray}\label{ec-CCNF-summary}
a \leftrightarrow \bigwedge^{n}_{i=1} \left( t \rightarrow \bigvee^{m}_{j=1} c^{j}_{i} \right), 
\end{eqnarray} 
where the variables  $c^{j}_{i}$ may contain negation but no other connective.
Let $D$ be the set of all possible summaries of this kind.
\item
If $\alpha$ and $\alpha_{ci}$ are defined by (\ref{ec-CCBL-gramar}), then  $\alpha \Leftrightarrow \alpha_{ci}$. Hence, it is true that a deductive discourse about $\alpha_{ci}$ is also a deductive discourse about $\alpha$; 
\end{itemize}
\end{definition}

%
%
%

\subsection{Inductive (Hilbertian) and deductive (Gentzen) discourse in CCBL}
\label{S3-4} \vspace{-4pt}

This section aims to show the analogy between deductive/inductive thinking and deductive/inductive logical computing in $CCBL$. 

\begin{figure*}[!ht]
\centering
\includegraphics[width=6in]{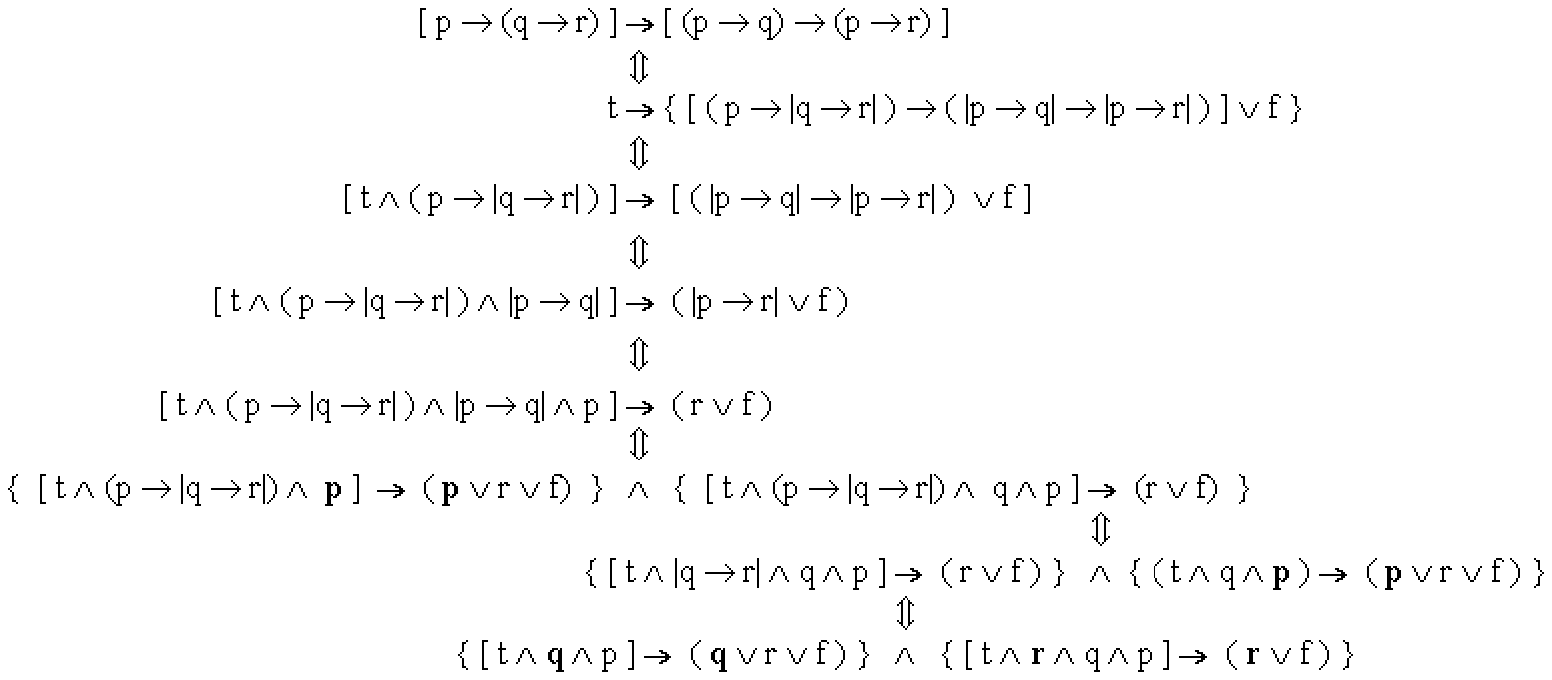}
\caption{A deductive discourse for one of Lukasiewicz's axioms. From top to buttom, the following expansion rules (valid arguments) are applied: 
\ref{ec-any-formula-is-an-implication4}, 
\ref{ec-varguments-2-6}, 
\ref{ec-varguments-2-6}, 
\ref{ec-varguments-2-6}, 
\ref{ec-varguments-2-5}, 
\ref{ec-varguments-2-5}, 
\ref{ec-varguments-2-5}.
Bold-face is used to mark instances of $TND$.}
\label{fig-deductive-discourse}
\end{figure*}

Let us consider an axiom of Lukasiewicz's formalization of propositional binary logic:
\begin{eqnarray}\label{ec-LBL-axiom-2}
\alpha = \{ [ p \rightarrow (q \rightarrow r) ] \rightarrow [ (p \rightarrow q) \rightarrow (p \rightarrow r) ] \},
\end{eqnarray}
Human/Computational deductive discourse about $\alpha$ try to equivalate $\alpha$ with (to `decompose' $\alpha$ as) a conjunction of simpler formulae with reduced structural complexity, up to the point where any component of this conjunction can  be easily proved, or tested, or reduced to a true hypothesis. An example is given in Fig.\ref{fig-deductive-discourse}. The deductive discourse presented there is not a \textit{full deductive discourse} because it still contains an expandable (a non-closed) leaf: 
$$\{ [ t \wedge (p \rightarrow |q \rightarrow r|) \wedge {\bf p} ] {\bf \rightarrow}  ({\bf p} \vee r \vee f) \}$$ 
on which the expansion rule (\ref{ec-varguments-2-5}) is still applicable. All of the other leaves are closed (non-expandable). All leaves are instances of $TND$ axiom (see formulae \ref{ec-the-axiom1}-\ref{ec-the-axiom3}). 

\begin{definition} \label{dfn_erdd}
Elements related to the deductive discourse in $CCBL$ (N. Popescu-Bodorin):
\begin{itemize}
\item
A node within a deductive discourse is  \textit{a terminal node / a closed node/ a closed leaf}  if the expansion rules of $\mathcal{A}_2$ are no longer applicable for that node, or else, it is an \textit{expandable node} (a continuation/non-terminal point of the deductive discourse).
\item
A deductive discourse is \textit{full} if all of its leaves are closed.
\item
\textit{A deductive proof} in $\mathcal{L}$ (\textit{a demonstration in CCBL}) is a full deductive discourse in $\mathcal{L}$ in which all leaves are instances of $TND$.
\item
A formula $\alpha$ (within $L$ or $\mathcal{L}$) is said to be \textit{deductively provable} (or, simply \textit{demonstrable} in $CCBL$) if there is a deductive proof for $\alpha$. 
\item
A \textit{deconstruction} in $\mathcal{L}$ is a full deductive discourse in $\mathcal{L}$  in which some leaves are logical but mutually irreconcilable, or some leaves are non-logical. 
\item
An \textit{illegal syntax} or a \textit{logical nonsense} (in $L$, $\mathcal{L}$, $LBL$, or $CCBL$) is a string whose full deductive discourse is a deconstruction.
\item
A contextual truth (in $L$, $\mathcal{L}$, $LBL$, or $CCBL$) is a well-formed formula whose full deductive discourse is neither a deductive proof, nor a deconstruction.
\end{itemize}
\end{definition}

At a first glance, the expression `\textit{closed leaf}' used above does not seem to be necessary, but in fact, it is mandatory, because in any partial expansion of a tree there are always current leaves. Some of them are expandable (becoming non-terminal nodes), or else, the expansion is full, not partial. The quality of being closed is not a geographic/pictographic attribute, but a semantic one.

A deductive proof of a formula is a failed deconstruction attempted for that formula which, in the end, instead of being deconstructed, proves to be a conjunction of true formulae (axiom instances). In a genetic view, a deductive proof of a formula shows that the formula carries only the logical genes of the truth, while a successful deconstruction reveals irreconcilable and contradictory meanings (logical genes of the false) and possibly non-logical genes of the given `formula' (see the \textit{Liar Paradox} example in section \ref{S3-9}). But generally speaking, the deductive discourse is neutral: it could be a deductive proof, or a successful logical deconstruction, or it could be none of them. The deductive discourses of the contextual truths fall into the latter category: their genes (the leaves) are reconcilable under certain hypotheses. 

A demonstration in Hilbertian acception is given by what we call a \textit{formal proof}, i.e. an inductive way of increasing semantic and syntactic complexity starting with some instances of the axioms, ending with the given formula and using a set of inference rules (valid arguments). In a way, a formal proof is the opposite of deduction. 

Since the valid arguments within $\mathcal{A}_{2}$ are equivalences, nothing could prevent us from reading them from the right to the left (as collapsing rules instead of expansion rules). Therefore, we can also consider $CCBL$ to be an inductive (Hilbertian) formal theory. Fortunately, the inductive formal proof can be parsed as a buttom-to-top traversal of the deductive proof. Consequently, in $CCBL$, inductive and deductive discourses of a formula perfectly match each other. This is why the distinction between the terms `deductively provable' and `inductively provable' is no longer necessary here.

\subsection{Completeness and Soundness of CCBL}
\label{S3-5} \vspace{-4pt}

The deductive discourse presented in Fig.\ref{fig-deductive-discourse} can be easily expanded to a deductive proof. Modus Ponens and Lukasiewicz's axioms (\ref{ec-LBL-axiom-1}-\ref{ec-LBL-axiom-3}) can be deductively proved by following the expansion rules of $\mathcal{A}_{2}$. Hence,  Lukasiewicz's formal theory of binary propositional logic is a sub-theory of $CCBL$, and therefore (see  formula \ref{ec-CCBL<SCBL<LBL}):
\begin{eqnarray}\label{ec-LBL<CCBL<SCBL<LBL}
LBL \subset CCBL \subset SCBL \subset LBL.
\end{eqnarray}

Hence, $CCBL$ and $SCBL$ theories are sound and complete. 

\subsection{CNF-ization}
\label{S3-6} \vspace{-4pt}

\begin{definition} A Cognitive Conjunctive Normal Form (CCNF) is a conjunction of cognitive implications:
\begin{eqnarray}\label{ec-CCNF}
\bigwedge^{n}_{i=1} \left( t \rightarrow \bigvee^{m}_{j=1} c^{j}_{i} \right), 
\end{eqnarray}
where the variables  $c^{j}_{i}$ may contain negation but no other connective.
\end{definition}

A full deductive discourse of a well-formed formula of $CCBL$ describe a conversion of that formula into a $CCNF$, i.e. a $CNF$-ization procedure.

$D$ is the set of all equivalences (\ref{ec-CCNF-summary}) between formulae within $FORM$ and their corresponding $CNF$. Any element of $D$ is the summary of a full deductive discourse.

\subsection{Semantic Closure of Propositional Binary Logic}
\label{S3-7} \vspace{-4pt}

Ultimately, binary logic is a study of two-element Boolean algebra and also a study of all the possible worlds which are compatible with logical formalization - a process of abstraction in which the facts and/or the rules within a universe are represented as elements of a \textit{logical formal language}, namely: logical constants, logical variables and logical formulae. In this context, one may be led to believe that  a logical variable and a logical formula are two different things. But, in fact, it is shown below that they aren't.

Let us recap what we have so far: $B=\{0,1\}$ - the set of binary truth values, $V$ - a set of truth-functional variables, $FORM$ - the set of well-formed formulae, $CI \subset FORM$ - the set of cognitive implications, $D$ - the set containing the summaries of the deductive discourses. But, what exactly is $V$? $V$ is a collection of symbols (labels) representing all kind of things which are true or false, but never simultaneously true and false. Therfore, in $CCBL$: $FORM \subset V$ and $D\subset V$, or in other words, $CCBL$ is the natural unified universe of discourse about binary truth values, truth-functional variables, logical formulae, deductive/inductive discourses and $CNF$-ization.

In other words, the vocabulary of propositional binary logic is rich enough to express its meanings. Since the vocabulary of $L$ contains all the elements of its meta-language $\mathcal{L}$, it is clear now that $CCBL$ formalization is nothing but an instrument meant to make obvious the semantic richness of binary logic and the structural complexity hidden behind the definition of $V$. We started by thinking that $V$ was a simple set of propositional variable and, at the end, we saw that the ways of reasoning about elements of $V$ belong in $V$. This is why we said that the (vocabulary of) \textit{propositional binary logic is semantically closed} (or self-described). Hence, there is no need to create/use a markup language for  describing the meanings of  propositional binary logic. 

\subsection{The fundamental property of binary logic vocabulary}
\label{S3-8} \vspace{-4pt}

\begin{figure}[!t]
\centering
\includegraphics[width=2.75in]{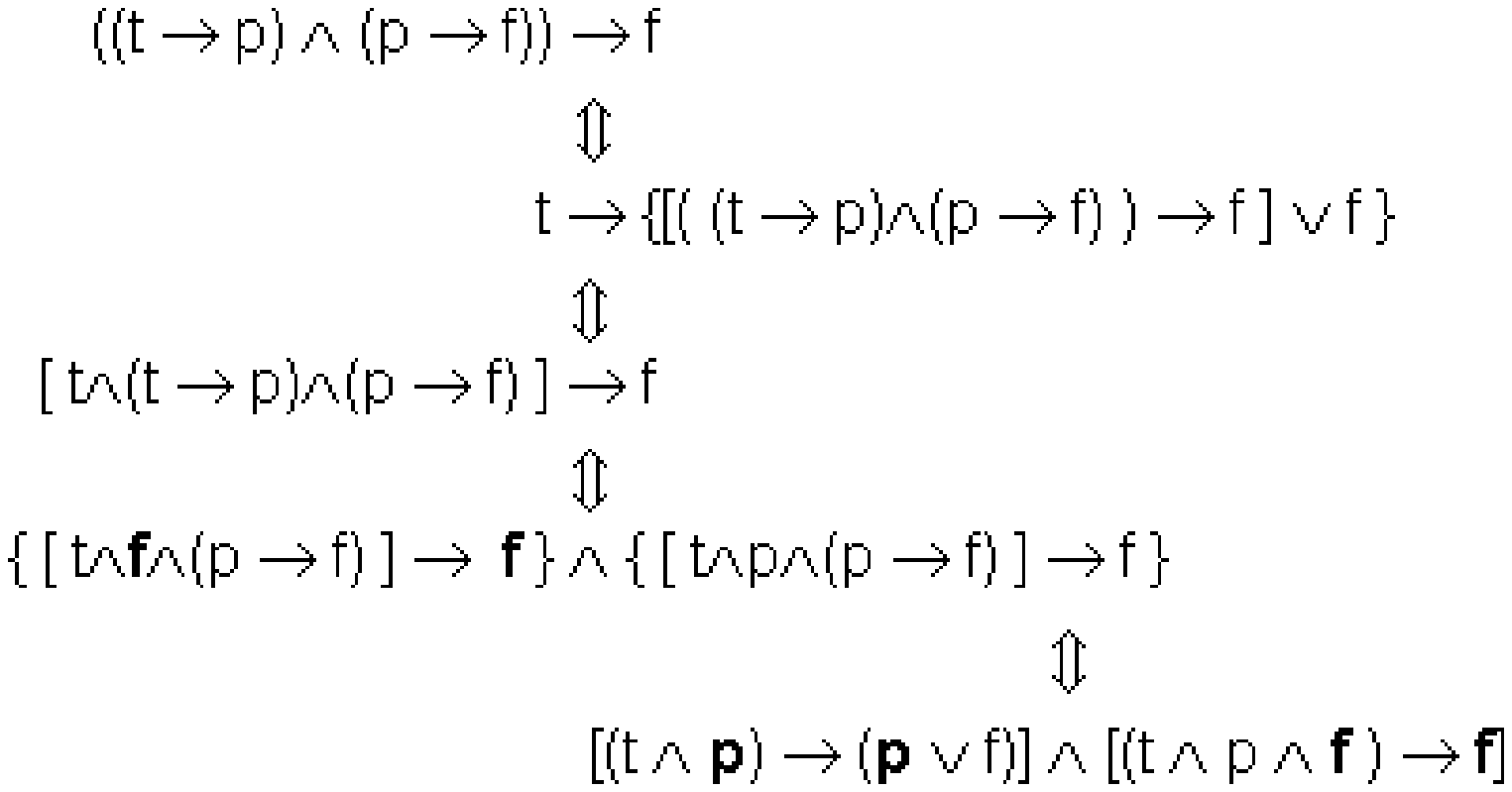}
\vspace{-10pt}
\caption{A deductive discourse for formula  (\ref{ec-fpblvocab-5})}
\label{fig-fpblvocab}
\end{figure}

If $p$ belongs to the vocabulary of propositional binary logic ($p \in V$ ), the following formulae are deductively provable in $CCBL$:
\begin{eqnarray}\label{ec-fpblvocab-1}
t \rightarrow [(t \rightarrow p) \rightarrow (\neg(p \rightarrow f))]
\end{eqnarray}
\begin{eqnarray}\label{ec-fpblvocab-2}
t \rightarrow [ (\neg(p \rightarrow f)) \rightarrow (t \rightarrow p) ]
\end{eqnarray}
The proof is that both formulae (\ref{ec-fpblvocab-1}),(\ref{ec-fpblvocab-2}) admit a deductive proof and the consequence is that, for any logical variable $p$, the following formula holds true: 
\begin{eqnarray}\label{ec-fpblvocab-3}
(t \rightarrow p) \leftrightarrow \neg(p \rightarrow f),
\end{eqnarray}
or, in other words, for any $p \in V$, the following formula is false:
\begin{eqnarray}\label{ec-fpblvocab-4}
(t \rightarrow p) \wedge (p \rightarrow f)
\end{eqnarray} 
Of course, the fact that formula  (\ref{ec-fpblvocab-4}) is false can be proved directly in $CCBL$. Fig.\ref{fig-fpblvocab} shows a deductive discourse (easily expandable to a deductive proof) of $CCBL$ formula: 
\begin{eqnarray}\label{ec-fpblvocab-5}
((t \rightarrow p) \wedge (p \rightarrow f)) \rightarrow f
\end{eqnarray} 

\vspace{-10pt}
\subsection{There is no Liar Paradox in CCBL!}
\label{S3-9} \vspace{-4pt}

One might believe that the \textit{Liar Paradox} gives an example of a well-formed formula of propositional binary logic which does not have a truth value. This is completely wrong. Let us investigate the existence of a true propositional variable $p \in V$ which says about itself that it is false. Since $CCBL$ is a complete and sound theory, if such a propositional variable exists, then it should be \textit{reachable} through a proof or a resolution.

Let us assume that there is such a propositional variable $p \in V$ so that it is true and says about itself that it is false. Let us translate from natural language - `it is true that $p$ is true and $p$ claims that it is false',   to the language of $CCBL$: 
\begin{eqnarray}\label{ec-LiarP}
t \rightarrow \{[(t \rightarrow p) \wedge (p \rightarrow f)] \vee f\}
\end{eqnarray}
The deductive way to disprove the existence of $p$ is to expand and to analyze the full deductive discourse of (\ref{ec-LiarP}). It should be clear that the full deductive discourse of  (\ref{ec-LiarP}) is a succesful deconstruction revealing that $p\not\in V$ , and more precisely that `\textit{formula}' (\ref{ec-LiarP}) carries genes of a paraconsistent logic \cite{parapara}. 

On the other hand, the existence of $p$ can be disproved in an inductive manner: by showing that assuming (\ref{ec-LiarP}) will rapidly lead to contradiction. Indeed, by applying the syllogism principle it is clear that, if (\ref{ec-LiarP}) is assumed to be true, the following four equivalent formulae should also be true: 
\begin{eqnarray}\label{ec-LiarPContrad}
t \rightarrow \{(t \rightarrow f) \vee f\}; t \rightarrow (t \rightarrow f); t \rightarrow f; f \vee f. 
\end{eqnarray}
But obviously, they aren't. Hence, the assumption that $p \in V$ is false and consequently, even if `formula' (\ref{ec-LiarP})  \textit{apparently seems to be well-formed} (as a pictogram), it is not semantically (hence, nor sintactically) well-formed because it contains a symbol extraneous to $V$. Consequently, `formula' (\ref{ec-LiarP}) is 
a string which confuses us with its pictographic mimetism, but nothing more. 

Hence, in propositional binary logic, the so called \textit{Liar Paradox} is just an abuse of formalization, a wrong attempt to apply logical principles beyond the boundaries of the vocabulary of classical propositional binary logic, a dangerous way to claim that classical binary logic (classical logical thinking) should be applicable to the vocabulary of a paraconsistent \cite{parapara} logical theory. 


In $CCBL$ theory, the so-called \textit{Liar Paradox} is totally and definitely deconstructed. An important consequence is that any formal theory in which the \textit{Liar Paradox} is a well-formed true formula doesn't contain a sufficient amount of binary logic. More precisely, if the \textit{Liar Paradox}, the syllogism principle, and formula $[(t\rightarrow \alpha)\leftrightarrow \alpha]$ are theorems within a theory, then that theory is inconsistent because, as shown above, in such a theory, the false formula $(t \rightarrow f)$ is provable.

\vspace{-1pt}
\section{Conclusion}
\label{S3} \vspace{-4pt}

Tarski held that the possibility of formulating paradoxically self-reference sentences belongs to semantically closed languages. He thought that a distinction between the language and the metalanguage is needed in order to avoid such faults. $CCBL$ formal theory gives a relevant counter-example. Humans can make mistakes but a deterministic machine can only do what is designed to do.



\vspace{-1pt}
\begin{thebibliography}{11}

\small
\vspace{-8pt}
\bibitem{gentzen}
M.~E.~Szabo, \textit{The collected papers of Gerhard Gentzen}, North-Holland Publishing Company, Amsterdam, 1970.

\vspace{-8pt}
\bibitem{prooftheory85}
J.~H.~Gallier, \textit{Logic for computer science: foundations of automatic theorem proving}, Harper \& Row Computer Science And Technology Series, New York, 1985.

\vspace{-8pt}
\bibitem{liar-paradox}
R.~M.~Sainsbury, \textit{Paradoxes}, Cambridge University Press, 2009.

\vspace{-8pt}
\bibitem{formal-language}
W.~J.~M.~Levelt,  \textit{Introduction to Theory of Formal Languages and Automata}, John Benjamins Publishing Company, 2008. 

\vspace{-8pt}
\bibitem{formal-theory}
H.~B.~Curry, \textit{A Theory of Formal Deducibility}, University of Notre Dame (Indiana), 1950.

\vspace{-8pt}
\bibitem{formal-proof}
R.~Bornat, \textit{Proof and Disproof in Formal Logic}, Oxford University Press, 2005.

\vspace{-8pt}
\bibitem{Lukasiewicz}
R.~Stansifer, \textit{Completeness of Propositional Logic as a Program}, Technical Report, Department of Computer Sciences, Florida Institute of Technology, March 2001.

\vspace{-8pt}
\bibitem{Hilbert}
D.~Hilbert, W.~Ackermann, \textit{Principles of Mathematical Logic}, Chelsea Publishing Company, 1950.

\vspace{-8pt}
\bibitem{ML}
P.~Blackburn, M.~de Rijke, Y.~Venema, \textit{Modal Logic}, Cambridge University Press, 2000.

\vspace{-8pt}
\bibitem{parapara}
J.~Woods, \textit{Paradox and paraconsistency: conflict resolution in the abstract sciences}, Cambridge University Press, 2003.

\end{thebibliography}
\end{document}